\documentclass{evn2002}
\begin{document}
   \title{2mm Wavelength VLBI of SiO Masers and AGN}

   \author{S. Doeleman\inst{1} 
	\and R. Phillips\inst{1}
	\and A. Rogers\inst{1}
	\and J. Attridge\inst{1}
	\and M. Titus\inst{1}
	\and D. Smythe\inst{1}
	\and R. Cappallo\inst{1}
	\and T. Buretta\inst{1}
	\and A. Whitney\inst{1}
	\and T. Krichbaum\inst{2}
	\and D. Graham\inst{2}
	\and W. Alef\inst{2}
	\and A. Polatidis\inst{2}
	\and U. Bach\inst{2}
	\and A. Witzel\inst{2}
	\and J. Zensus\inst{2}
	\and A. Greve\inst{3}
	\and M. Grewing\inst{3}
	\and R. Freund\inst{4}
	\and P. Strittmatter\inst{5}
	\and L. Ziurys\inst{5}
	\and T. Wilson\inst{5}
	\and H. Fagg\inst{5}
	\and G. Gay\inst{5}
          }

   \institute{MIT-Haystack Observatory, Off Route 40, Westford, MA, USA 01886 
         \and
             Max-Planck-Institut für Radioastronomie, Auf dem Hügel 69, D-53121 Bonn, Germany
	\and
		IRAM, Grenoble, 300 Rue de la Piscine, 38460 St. Martin d'Héres, France
	\and 
		NRAO, 949 North Cherry Avenue, Tucson, AZ, USA 85721
	\and
	Steward Observatory, University of Arizona, Tucson, AZ, USA 85721	
             }

   \abstract{In April 2002 an array of antennas operating at 129GHz successfully
detected VLBI fringes on both continuum AGN and SiO spectral line sources.
The 129GHz fringes on maser sources represent the highest frequency spectral line VLBI
detections to date.  The AGN 3C279 was detected on long baselines at both 129GHz (and 
at 147GHz, see Krichbaum et al in these proceedings) yielding fringe spacings of 
50-56 micro arc seconds,
an angular resolution record.  The array consisted of the University of
Arizona Kittpeak 12m antenna, the Heinrich Hertz 10m Telescope (HHT), and the
IRAM 30m dish on Pico Veleta.  

At 129GHz, a number of evolved stars and several young stellar objects exhibit
strong SiO maser emission in the v=1 J=3-2 transition.  Preliminary cross
power spectra of SiO masers around the red hypergiant VYCMa on the HHT-KittPeak 
baseline ($\sim$190km) are consistent with
multiple spatially separate maser spots associated with the star.  
Future observations will include continuum observations of the radio source 
at the Galactic Center,
SgrA*, and higher frequency maser lines including HCN and methanol.
   }

   \maketitle
%

\section{Introduction}

A compelling case can be made for extending the VLBI technique to high
frequencies.  First and foremost, angular resolution improves linearly with
frequency allowing more detailed imaging of sources.  This is the only avenue
available for better resolution given Earth-based arrays: VLBI at 3mm
wavelength already approaches 80$\mu$as fringe spacing which is very near the
limit imposed by the Earth's diameter.  Plasma effects which can mask small
scale structure at lower frequencies also decrease in severity as $\lambda^2$.
Faraday Rotation (and therefore depolarization) is a prime example, and studies
of jet formation and propagation in AGN, which depend on teasing out the fine
structure of B-fields, require high frequency VLBI.  Scatter broadening due to
ionized components of the ISM is another example, one which prevents imaging
the intrinsic structure of Sgr A*, the massive black hole candidate in our
Galaxy.  In addition, radio spectra of AGN generally turn over in the mm
wavelength range, so high frequency VLBI arrays can peer more deeply into AGN
cores that are Synchrotron Self Absorbed at lower frequencies.  And, finally,
by opening new spectral windows, previously inaccessible maser transitions
become available for VLBI study.

With many new large aperture antennas planned in the mm and sub-mm
bands, it is not so much a question of {\it if}, but of {\it when}
VLBI at frequencies higher than 86 GHz will deliver on the promise
outlined above.  Within the next decade, CARMA (a combination of the
current BIMA and OVRO arrays), ALMA and the LMT will combine with
existing mm wave dishes, including the IRAM 30m, to enable Earth-sized
high frequency VLBI arrays with enough sensitivity to image AGN at
useful dynamic ranges.  Current arrays are limited at high frequencies
in both sensitivity and baseline coverage, but they provide crucial
test beds for developing techniques and exploring future scientific
drivers.  

We report here on a successful VLBI experiment at 129GHz in April 2002
on a triangle of stations including the IRAM 30m (Spain), the Heinrich
Hertz Telescope (Mt. Graham, Arizona) and the University
of Arizona 12m (KittPeak, Arizona).  On the long baselines to IRAM,
3C279 was detected with fringe spacings of 56$\mu$as.  Such detections
are useful in estimating AGN core sizes, but this sparse array does
not permit true imaging.  The main targets of the 129GHz experiment
were not AGN but SiO J=3$\rightarrow$2 maser sources around both
evolved stars and the Orion star forming region.  Because different
spectral features of masers can correspond to distinct regions of
emission in the circumstellar environment, even sparse VLBI arrays can
often map the maser structure.  Here we describe the technical work at
both the KittPeak and HHT sites that enabled this experiment and
discuss the preliminary results and prospects for imaging.

\section{Technical Aspects}

   \begin{figure*}
   \centering
   \vspace{250pt}
   \includegraphics{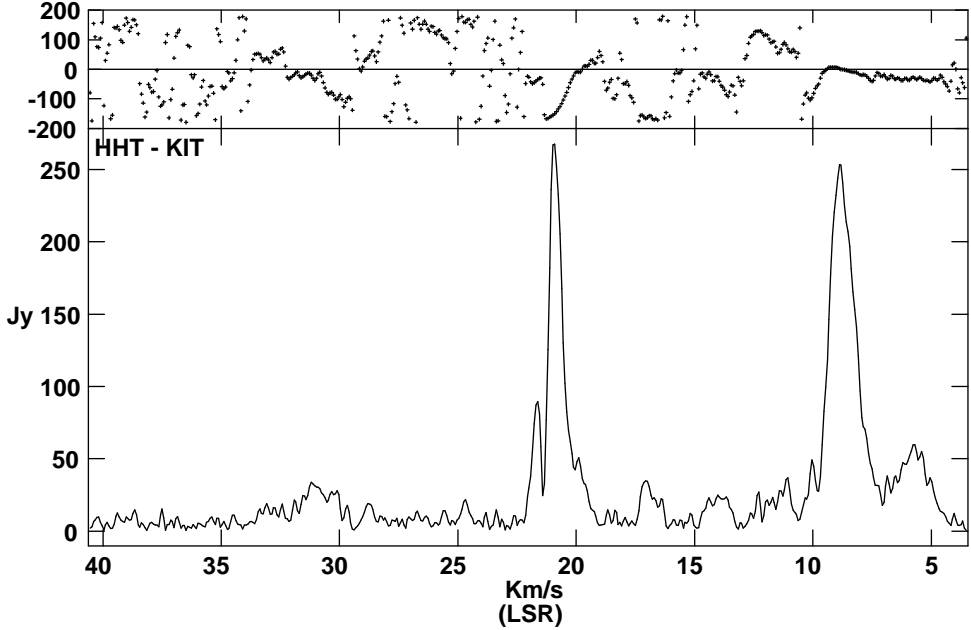}
   \caption{Cross power spectra of the v=1 J=3$\rightarrow$2 (129 GHz) SiO maser
toward the evolved star VYCMa.  Top panel shows interferometric phase,
bottom panel the amplitude.  Spectral features with distinct phases
are most likely spatially separated maser features in the
circumstellar envelope.  The data are calibrated in Jy assuming gains
at both HHT and KittPeak sites of 0.02K/Jy and Tsys of 300K and 400K
respectively.
            \label{fig:crosspower}
           }
    \end{figure*}

A number of technical tasks had to be accomplished to carry out the
2mm VLBI.  Key technical aspects are briefly described below.

A previous 230GHz VLBI attempt using the HHT revealed that the
Hydrogen maser used at that site may not have been sufficiently
stable.  This unit was shipped back to the CfA in Cambridge, MA. where
it was found to have a stability characterized by: $\sigma_y(\tau) =
4\times10^{-13}\tau^{-1} + 8\times10^{-14}\tau^{-0.5}$.  The coherence
loss at this level of maser stability can be expressed as
$\mathrm{Loss}=1-\exp(-\sigma(\tau)^2/2)$ where
$\sigma(\tau)=2\pi\nu\tau\sigma_y(\tau)$.  So for $\tau=10$ seconds,
the coherence loss would be $\sim36$\%, an unacceptably high value.  A
newer H-maser was refurbished, tested and shipped out to the HHT in
preparation for observations at 129GHz and 147GHz (the 147 GHz
observations are covered elsewhere in these proceedings).  

To search for VLBI fringes, the exact position of the HHT was
required.  A geodetic differential GPS antenna was mounted on the back
side of the HHT subreflector and the position refined over a 24 hour
period.  The same measurement was performed at the KittPeak 12m whose
VLBI position was known, and the resulting position difference yielded
the HHT position to within $\sim10$cm.  

The decision to attempt the VLBI in the 2mm band as opposed to the 1mm
band was motivated by the decrease in sensitivity to maser stability,
atmospheric turbulence and instrumental phase noise.  Accordingly, a
2mm receiver had to be constructed for use at the HHT - the KittPeak
12m and IRAM 30m were already so equipped. 

VLBI backends including data acquisition racks and magnetic tape
recorders were shipped to both HHT and KittPeak sites for this
experiment.  Both sites used Mark4 VLBI formatters and recorded total
bit rates of 224Mb/s using 2-bit samples to increase sensitivity on
spectral line sources.

Broadband $\frac{1}{4}\lambda$ plates were fabricated at IRAM for use
at the two Arizona sites.  These plates were sufficiently broadband
that they could be used at both 129 GHz and 147GHz to convert LCP to 
linear polarization.

Local oscillators at all sites were tested by injecting a tone
directly into the receiver feeds and mixing down to IF frequencies.
Comparison of this tone with one derived from the H-maser frequency
standard can expose potential coherence losses due to LO impurity.  At
KittPeak, for example, a high frequency ($\nu>1\mathrm{MHz}$) noise
pedestal on the LO caused $\sim10$\% of the signal power to be moved
outside the VLBI fringe rate window.

\section{SiO Masers}

Silicon Monoxide (SiO) masers, excited by a combination of radiative and
collisional pumps, arise in some atmospheres of both evolved stars and YSOs.
Masers are observed in many ro-vibrational transitions of this simple rotor
molecule, giving rise to very compact and very bright features that trace long
gain paths through stellar environments.  The excitation energy of the v=1 SiO
vibrational state is 1800K which places these masers, due to pumping
considerations, very close to the stellar photosphere.  VLBI observations place
the masers within $\sim4$ stellar radii of evolved stars (Diamond et al
\cite{diamond}, Doeleman et al \cite{doeleman98}, Phillips et al
\cite{phillips}).  SiO masers can be excellent dynamical probes providing
evidence for rotation in evolved stars (Boboltz et al \cite{boboltz}, Hollis et
al \cite{hollis}) and for outflow from protostars (Greenhill et al
\cite{greenhill}, Doeleman et al \cite{doeleman99}).  In most cases, VLBI of SiO
masers provides the most detailed pictures of the envelopes around these stellar
objects which are opaque in the optical and IR.

Recent VLBI studies reveal that the relationship between different SiO maser
transitions around individual objects is complex.  Towards the Red Giant R
Cassiopeiae, for example, the v=1 J=1$\rightarrow$0 (43 GHz) and
J=2$\rightarrow$1 (86 GHz) masers appear to be largely cospatial (Phillips et
al, these proceedings).  In the Orion-BN/KL region, however, these same
transitions are offset from each other and likely occur in different regions of
a shocked bipolar outflow (Doeleman et al \cite{doeleman02}).  The HHT-KittPeak 
baseline is unique in its capacity to observe SiO masers in the
J=3$\rightarrow$2 (129 GHz), J=4$\rightarrow$3 (172 GHz) and J=5$\rightarrow$4
(215 GHz) SiO transitions with angular resolutions comparable to known maser
structures.  Given the results on lower frequency maser lines, VLBI exploration
of SiO masers in the 1-2mm band will likely reveal unforeseen relationships among
newly imaged transitions.

In contrast with the case of mapping continuum sources, maser emission can be
imaged with a single baseline, and the peculiarities of doing so are well
understood (Doeleman et al 1998).  It is essential that an isolated spectral
feature be found that is very closely approximated by a point source.  To do
this, one must examine the calibrated visibility amplitudes at each frequency
channel.  Once a point-like feature is found, the rest of the data set is
spectrally phase referenced to this feature and mapping can proceed.

During the April 2002 observations, we observed a total of six SiO
maser sources selected for their flux density.  These sources are
highly variable with the SiO maser maximum typically lagging the
IR/Optical maximum by 0.25 of a stellar cycle.  The sources were:
Orion-BN/KL, WHya, VYCMa, RLeo, RCas, $\chi$Cyg.

The data reduction at this point is preliminary and we have only detected one
source, VYCMa, so far.  The cross-power spectrum of VYCMa
(Figure~\ref{fig:crosspower}) reveals strong compact emission at multiple
velocities.  The associated phases (top panel) show that after fringe rate
correction, there are significant phase offsets between spectral features.
These data have not yet been corrected for phase delay across the passband, but
it is likely, given the phase signatures, that we are seeing individual maser
features which are spatially distinct.  It remains to be seen if we can isolate
a point-like feature and produce a real map of the emission.  The ability,
though, of the Mark4 correlator to produce 512 point cross power spectra
increases the chances that such a component can be found.

   \begin{figure}
   \centering
   \vspace{225pt}
   \includegraphics{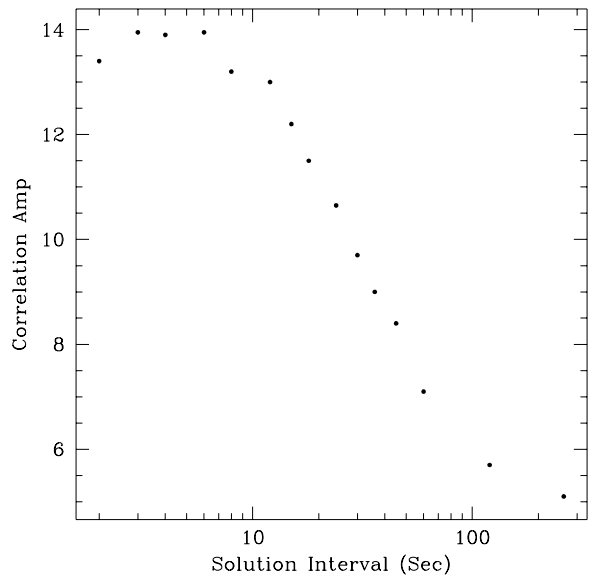}
      \caption{Coherence of the HHT-KittPeak baseline was determined
for a sample scan by plotting the vector averaged amplitude of the spectral feature at
8.5 km/s as a function of fringe solution interval.  The plateau below
intervals of $\sim8$ seconds sets the coherence time.
         \label{fig:coherence}
         }
   \end{figure}
%

The SiO masers in VYCMa are bright enough that fringe rate solutions can be
found on time scales below 3 seconds.  This allows us to empirically determine
the coherence time of the HHT-KittPeak baseline at 129GHz by comparing the
strength of spectral features at different solution intervals.  Figure
~\ref{fig:coherence} shows the resulting plot in which a clear ``plateau" forms
at solution intervals below 8 seconds.  This occurs because the solution
interval falls below the coherence time of the interferometer.  If we take the
coherence time to be that solution interval at which the amplitude of the
spectrum falls to 95\% of this ``plateau" value, then this baseline has a
$\sim10$sec coherence time.  This is almost certainly due to atmospheric
turbulence.

\section{AGN and Sgr A*}

The AGN 3C279 served as a calibrator for the spectral line targets, but was
also detected at 129GHz on intercontinental baselines from Arizona to the IRAM
30m in Spain.  If we assume that 3C279 must be approximately half the size of
the 56$\mu$as fringe spacing, then its linear size must be less than 0.2 pc
($H_\circ$=100km/s/Mpc, $q_\circ$=0.5, z=0.538).  This size corresponds to 2000
Shwarzschild Radii of a $10^9M_\odot$ Black Hole.

By segmenting and incoherently averaging the 3C279 visibilities, we find that
during these observations the coherence times on the HHT-KTPK, HHT-IRAM,
KTPK-IRAM baselines were 2 seconds, 8 seconds and 3 seconds respectively.
Comparison with the $\sim10$ seconds coherence time in Fig 2 implies that poor
weather at the Kittpeak site is responsible.  After correcting for the effects
of short coherence time, we find correlation amplitude values for 3C279 of
Amp(KTPK-IRAM)=3.5x$10^{-4}$, Amp(KTPK-HHT)=5.2x$10^{-4}$ and
Amp(HHT-IRAM)=5.2x$10^{-4}$.  

We attempted to calibrate these data assuming SEFD(HHT)=15000Jy,
SEFD(IRAM)=2600Jy.  By further assuming that the correlated flux density on both
Arizona-Spain baselines must be identical, we find that SEFD(KTPK)=33000Jy.
This calibration gives the correlated flux densities in Table~\ref{tab:corrflux}.

During the observations, the total flux density of 3C279 was measured to be
$\sim21Jy$ at 129GHz.  If we assume that most of the 129GHz flux is compact on
$\sim$2mas scales, then the measured short baseline
${\mathrm{S}_{\mathrm{corr}}}$ is unreasonable and may be due to any of a
number of calibration uncertainties: 1. there is some source of undiagnosed
phase noise in the signal path; 2. coherence times are actually smaller than we
are measuring; 3. the calibration (Tsys) is faulty; or 4. some problem we do
not yet understand.  In any event, VLBI calibration at these frequencies
remains a difficult problem and will likely be solvable only with the inclusion
of more telescopes so that closure quantities can be used.  Even with the
calibration uncertainty, it appears that 3C279 is resolved on the long
baselines.

   \begin{figure}
   \centering
   \vspace{250pt}
   \includegraphics{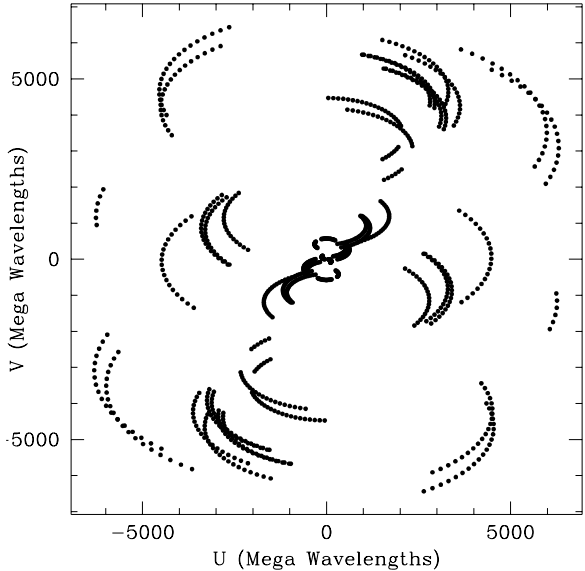}
      \caption{230GHz-VLBI baseline coverage on the Galactic Center source
Sgr A* on an array including: CARMA, LMT, ALMA, KittPeak, HHT, IRAM
30m, SEST, JCMT.
         \label{fig:sgrauv}
         }
   \end{figure}
%

   \begin{table}
      \caption[]{3C279 Calibrated Flux Densities.}
         \label{tab:corrflux}
     $$
         \begin{array}{p{0.25\linewidth}cc}
            \hline
            \noalign{\smallskip}
            Baseline      &  {\mathrm{S}_{\mathrm{corr}}} (Jy) & Baseline (M{\lambda}) \\
            \noalign{\smallskip}
            \hline
            \noalign{\smallskip}
            HHT-IRAM & 2.9 & 3600    \\
            KTPK-IRAM & 2.9 & 3670  \\
            HHT-KTPK & 11.6 & 73   \\
            \noalign{\smallskip}
            \hline
         \end{array}
     $$
   \end{table}

Sgr A*, the compact radio source at the Galactic Center is thought to mark the
position of a $2.5\times10^6M_\odot$ Black Hole.  While it is tantalizingly
close to Earth and holds the promise of helping us understand the physics of its
more massive extragalactic cousins, it is shrouded by an ionized plasma that broadens VLBI
images.  High frequency VLBI is the only means available to pierce the
scattering screen and image the intrinsic structure of Sgr A*.  Doeleman et al
(\cite{doeleman01}) have shown that it is possible to use closure quantities to
calibrate a high frequency (86 GHz) VLBI array and iteratively solve for
structural models of Sgr A*.  But this requires of order 6 telescopes to ensure
that enough phase and amplitude data can be recovered using closure.  At
frequencies above 86 GHz, current VLBI arrays become sparse and the sensitivity
of individual antenna are relatively low.  

Two factors will alter the future VLBI landscape and permit 1mm and 2mm
observations of Sgr A*.  In the near future, the Mark5 VLBI system (see A.
Whitney, these proceedings) will boost recording capability from the standard
256 Mb/s to over 1Gb/s thereby increasing baseline sensitivity by over a factor
of 2.  This enhancement puts Sgr A* within reach at 230GHz on today's longest
and most sensitive baselines.  Sgr A* has already been detected with 215GHz VLBI
(Krichbaum et al \cite{krichbaum}, Greve et al \cite{greve}) on moderate length
baselines, but it is difficult to make size estimates due to calibration
uncertainty.  

The second factor is that a number of large aperture mm wave dishes and arrays
are planned for construction over the next decade.  These sensitive dishes
combined with wideband recording should be able to definitively map the
structure of Sgr A*.  Figure~\ref{fig:sgrauv} shows the $(u,v)$ coverage
possible with such an array.

\begin{acknowledgements}
We thank all personnel at the participating observatories for their assistance.  We also thank
Robert Vessot and Dick Nicoll of the Harvard-CfA for their assistance with the HHT hydrogen maser.
\end{acknowledgements}


\begin{thebibliography}{}

   \bibitem[2000]{boboltz} Boboltz, D. \& Marvel, K.B. 2000, ApJ, 545, L149

   \bibitem[1994]{diamond} Diamond, P.J., Kemball, A.J., Junor, W., Zensus, A., 
    Benson, J., Dhawan, V. 1994, ApJ, 430, L61

   \bibitem[1998]{doeleman98} Doeleman, S., Lonsdale, C.J., Greenhill, L.J. 1998, ApJ, 
    494, 400

   \bibitem[1999]{doeleman99} Doeleman, S., Lonsdale, C.J., Pelkey, S. 1999,
    ApJ, 510, L55

   \bibitem[2001]{doeleman01} Doeleman, S., Shen, Z.-Q., Rogers, A.E.E., Bower, G.C., 
    Wright, M.C.H., Zhao, J.H., Backer, D.C., Crowley, J.W., Freund, R.W., Ho, P.T.P., 
    Lo, K.Y., Woody, D.P. 2001, AJ, 121, 2610 

   \bibitem[2002]{doeleman02} Doeleman, S., Lonsdale, C.J., Kondratko, P.,
    Predmore, C.R. 2002, in proceedings of the IAU Symp. 206 "Cosmic MASERs: from 
    protostars to blackholes", 2002, Brazil., Eds. Victor Migenes and Mark J. Reid., 
    vol 206, p282.

   \bibitem[1998]{greenhill} Greenhill, L.J., Gwinn, C.R., Schwartz, C., Moran, J.M., 
    Diamond, P.J. 1998, Nature, 396, 650.

   \bibitem[1995]{greve} Greve, A., Torres, M., Wink, J.E., Grewing, M., Wild, W., Alcolea, J., 
    Barcia, A., Colomer, F., de Vincente, P., Gomez-Gonzalez, J., Lopez-Fernandez, I., 
    Graham, D.A., Krichbaum, T.P., Schwartz, R., Standke, K.J., Witzel, A., Baudry, A. 1995, 
    A\&A, 299, L33

   \bibitem[2001]{hollis} Hollis, J.M., Boboltz, D.A., Pedelty, J.A., White, S.M., 
    Forster, J.R. 2001, ApJ, 559, L37

   \bibitem[1998]{krichbaum} Krichbaum, T.P., Graham, D.A., Witzel, A., Greve, A., 
    Wink, J. E., Grewing, M., Colomer, F., de Vicente, P., Gomez-Gonzalez, J., Baudry, A., 
    Zensus, J.A. 1998, A\&A, 335, L106
 
   \bibitem[2001]{phillips} Phillips, R.B., Sivakoff, G.R., Lonsdale, C.J., Doeleman, S.
    2001, AJ, 122, 2679

\end{thebibliography}
\end{document}